\documentclass[showpacs,preprintnumbers,amsmath,amssymb]{revtex4}
\usepackage{amsmath}
\usepackage{graphicx}
\usepackage{subfig}
\usepackage{hyperref}
\usepackage{amssymb}
\usepackage[english]{babel}
\usepackage{epsfig}
\usepackage{wasysym}
\usepackage{graphicx}
\usepackage{indentfirst}
\usepackage{amsmath}
\usepackage{amsfonts}
\usepackage{amssymb}

\begin{document}

\newcommand{\vecbo}[1]{\mbox{\boldmath $#1$}}
\newtheorem{theorem}{Theorem}
\newtheorem{acknowledgement}[theorem]{Acknowledgement}
\newtheorem{algorithm}[theorem]{Algorithm}
\newtheorem{axiom}[theorem]{Axiom}
\newtheorem{claim}[theorem]{Claim}
\newtheorem{conclusion}[theorem]{Conclusion}
\newtheorem{condition}[theorem]{Condition}
\newtheorem{conjecture}[theorem]{Conjecture}
\newtheorem{corollary}[theorem]{Corollary}
\newtheorem{criterion}[theorem]{Criterion}
\newtheorem{definition}[theorem]{Definition}
\newtheorem{example}[theorem]{Example}
\newtheorem{exercise}[theorem]{Exercise}
\newtheorem{lemma}[theorem]{Lemma}
\newtheorem{notation}[theorem]{Notation}
\newtheorem{problem}[theorem]{Problem}
\newtheorem{proposition}[theorem]{Proposition}
\newtheorem{remark}[theorem]{Remark}
\newtheorem{solution}[theorem]{Solution}
\newtheorem{summary}[theorem]{Summary}
\newenvironment{proof}[1][Proof]{\textbf{#1.} }{\ \rule{0.5em}{0.5em}}
\hypersetup{colorlinks,citecolor=green,filecolor=magenta,linkcolor=red,urlcolor=cyan,pdftex}
\newcommand{\be}{\begin{equation}}
\newcommand{\ee}{\end{equation}}
\newcommand{\bea}{\begin{eqnarray}}
\newcommand{\eea}{\end{eqnarray}}
\newcommand{\beaa}{\begin{eqnarray*}}
\newcommand{\eeaa}{\end{eqnarray*}}
\newcommand{\Lhat}{\widehat{\mathcal{L}}}
\newcommand{\nn}{\nonumber \\}
\newcommand{\e}{{\rm e}}


\title{Thermodynamics of  phantom Reissner-Nordstrom-AdS black hole}

\author{ Deborah F. Jardim $^{(a)}$\footnote{E-mail
address: dfjardim@gmail.com}
, M. E. Rodrigues $^{(b)}$\footnote{E-mail
address: esialg@gmail.com} and M. J. S. Houndjo $^{(c)}$\footnote{E-mail address:
sthoundjo@yahoo.fr} }
\medskip
\affiliation{$^{(a)}$ Universidade Federal dos Vales do Jequitinhonha e Mucuri, ICTM\\
Rua do Cruzeiro, 01, Jardim S\~{a}o Paulo, CEP39803-371 - Teofilo Otoni, MG - Brazil}
\affiliation{$^{(b)}$ Centro de Ci\^{e}ncias
Exatas - Departamento de F\'{\i}sica - Universidade Federal do Esp\'{\i}rito Santo\\
Av. Fernando Ferrari s/n - Campus de Goiabeiras - CEP29075-910 -
Vit\'{o}ria/ES - Brazil}
\affiliation{$^{(c)}$ Departamento de Ci\^{e}ncias Naturais - CEUNES - Universidade Federal do Esp\'irito Santo - 
CEP 29933-415 - S\~ao Mateus - ES - Brazil}
\affiliation{$^{(d)}$Institut de Math\'{e}matiques et de Sciences Physiques (IMSP) - 01 BP 613 Porto-Novo, B\'{e}nin}

\begin{abstract}
We obtain a new solution of the Einstein-anti-Maxwell theory with cosmological constant, called anti-Reissner-Nordstrom-(A)de Sitter (anti-RN-(A)dS) solution. The basic properties of this solution is reviewed. Its thermodynamics is consistently  established,  with the extreme cases and phase transitions,  where the  analysis is  performed  through two methods, the usual one and that of Geometrothermodynamics . The  Geometrothermodynamics analysis  does not provide  a result in agreement with the usual method or by the specific heat. We establish local and global thermodynamic stabilities of anti-RN-AdS solution through the specific heat and the canonical and grand-canonical ensembles.
\end{abstract}
\maketitle 
\section{Introduction}
\hspace{0,6cm} 
The discovery made by Hawking \cite{hawking1}, that black holes radiate particles by gravitational interaction near the event horizon, being interpreted as the temperature of the black hole, allowed numerous conjectures for various thermodynamic systems for a variety  of black holes solutions in General Relativity, and other modifications of gravity. This analysis of thermodynamic systems of black holes, as well as the usual thermodynamics, have their base in four postulates, called zero law and three other laws, by analogy to the usual system of thermodynamics \cite{bardeen,lousto}. This becomes quite interesting, since each new black hole solution can possess a different thermodynamic system, but which is ruled by laws, or fundamental principles, which are often up to universal,  as in usual thermodynamics. The resulting properties of these principles can give us good indications on the local or global stability of these systems, so a good criterion for finding a solution that can be physically acceptable or not.
\par
The thermodynamic properties of a black hole solution can be analysed in many ways. One of them is the usual study made by Davies \cite{davies}. But there are other methods well established both physically and mathematically, which are similarly used to study the thermodynamic properties, such as that of the Geometrothermodynamics \cite{quevedo} and that of Hamiltonian thermodynamics \cite{louko}. Our main goal here is to use the first two methods in parallel in order to facilitate comparison of the results and their analysis. Because these methods can give us a good notion of the thermodynamic stability of the system, we will study the properties of a new black hole solution obtained for the first time in this work. This solution has a unique feature which may seem strange at a first sight. It arises from a coupling of a field of spin $1$, with gravitation, which may be the usual Maxwell one, or with a contribution of negative energy density, called phantom solution.
\par 
Before beginning the analysis of this new class of phantom black holes, we will briefly present our interest in obtaining and studying such exotic solutions. With the discovery of the acceleration of the universe, various observational programs of the study of the evolution of our universe were deployed, including the relationship of the magnitude-versus-redshift type Ia supernovae and the spectrum of the anisotropy  of the cosmic microwave background. These programs promote an accelerated expansion of our universe, which should be dominated by an exotic fluid and should have a negative pressure. Moreover, these observations show that this fluid can be phantom, i.e., with the contribution of negative energy density \cite{hannestad,r1}.
\par
As the interest in obtaining these classes has increased, we also found ourselves wanting to analyse a specific phantom model. We can mention here some recent results in the literature, such as the wormhole solutions \cite{kirill1}, the black hole solutions of Einstein-Maxwell-Dilaton theory, \cite{gibbons}, the higher-dimensional black holes by Gao and Zhang \cite{gao}, and the higher-dimensional black branes by Grojean et al \cite{grojean}. Analyses were also made in algebraic structures of this type of phantom system, as the case of the algebra generated by metrics depending on two temporal coordinates, with $ D\geq 5$, which provides phantom fields in $ 4D$, fulfilled by Hull \cite{hull}, and Sigma models by Clement et al \cite{gerard2}. Here, we will obtain and study the thermodynamic properties of a solution arising from the coupling of Einstein-Hilbert action with a field of spin $1$ and a cosmological constant.
\par
One way to analyse the thermodynamics of black holes that has been widely used is the geometry, for which a thermodynamic phase space is defined, which is metric, and from differential geometry  all of the information of the properties of the system is removed. These methods provide us, through the curvature scalar of the space phase, if the system presents thermodynamic interaction, showing were the localisation of the point of the limit of extreme black, and where and when the system passes through a phase transition, and finally, if the system is  locally or globally stable or not. These methods have originated in the work of Rao \cite{rao}, which was subsequently developed by other authors \cite{amari}. Other works that have gained attention in the analysis of the thermodynamics of black holes by differential geometry were those of Weinhold \cite{weinhold} and Ruppeiner \cite{ruppeiner,r2}. We use here the so-called  Geometrothermodynamics method, however, comparing with the more commonly method,  used  for the  verification and interpretation of the obtained results. Geometrothermodynamics appears to be a method which is equivalent or superior in some ways, as for example, due to the invariance of Legendre transformations and to the consistency of the results in the various selected ensembles. This superiority can also be evidenced in the description through only one mathematical structure by this method,  in contrast to the many steps that the usual procedure uses to study the points of second order phase transition, the local and global stabilities, the fluctuations of second order, the correlations and the representations in various ensembles.  Recently,  this method has gained some attention by the scientific community \cite{quevedo1}-\cite{r3}. This case, which we will study, is a good test for this method because some difficulties already appeared in the utilization of this method applied to phantom solutions, showing some weakness, or even limitation of the method \cite{zui}.
\par
This paper is organized as follows. In section $2$, we present a summary of what to obtain a new class of solutions studied, arising from the Einstein-(anti)Maxwell theory, and how to obtain their thermodynamic quantities. In section 3, we summarize the method of Geometrothermodynamics. In section 4, we apply the method of Geometrothermodynamics for a new black hole solution and the Reissner-Nordstrom-AdS solution. In section 5, the conclusions and perspectives are presented.


\section{\large The field equations and the solutions of phantom Reissner-Nordstrom-(A)dS}

The action of this theory is given by 
\begin{eqnarray}
S=\int d^4 x\sqrt{-g}\left[R+2\eta F_{\mu\nu}F^{\mu\nu}+2\Lambda\right]\label{action}\,,
\end{eqnarray}
where the first term characterizes the Einstein-Hilbert action, the second is the coupling with the Maxwell field  ($\eta=1$), or a phantom field of spin $1$ ($\eta=-1$), and finally the third term is the coupling with the cosmological constant, which can behave as  $\Lambda>0$ (dS) or $\Lambda<0$ (AdS). The nomenclature known as phantom is due to the fact that the energy density of the field of spin  $1$ is negative, for $\eta=-1$.
\par
The conventions adopted here are with the description of the metric whose the signature is $+---$, the Ricci tensor defined as  $R_{\mu\nu}=\partial_{\alpha}\Gamma^{\alpha}_{\;\;\mu\nu}-\partial_{\nu}\Gamma^{\alpha}_{\;\;\alpha\mu}+\Gamma^{\lambda}_{\;\;\mu\nu}\Gamma^{\alpha}_{\;\;\alpha\lambda}-\Gamma^{\lambda}_{\;\;\alpha\mu}\Gamma^{\alpha}_{\;\;\lambda\nu}$, where $\Gamma^{\alpha}_{\;\;\mu\nu}$ is the connection of Levi-Civita, and the covariant derivative defined by $\nabla_{\mu}V_{\nu}=\partial_{\mu}V_{\nu}-\Gamma^{\lambda}_{\;\;\mu\nu}V_{\lambda}$. The Maxwell 2-form is given by $F=\frac{1}{2!}F_{\mu\nu}dx^{\mu}\wedge dx^{\nu}$, with the components $F_{\mu\nu}=\partial_{\mu}A_{\nu}-\partial_{\nu}A_{\mu}$, $A_{\mu}$ being the four-potential.
\par
Making the functional variation of the action  (\ref{action}), with respect to the fields  $A_{\mu}$ and  $g_{\mu\nu}$ (components of the spacetime metric), we get the following field equations \begin{eqnarray}\label{ec}
\left\{ \begin{array}{ll}
\nabla_{\mu}F^{\mu\nu}=0\;,\\
R_{\mu\nu}=2\eta\left(\frac{1}{4}g_{\mu\nu}F_{\alpha\beta}F^{\alpha\beta}-F_{\mu}^{\;\;\alpha}F_{\nu\alpha}\right)-\Lambda g_{\mu\nu}\;,
\end{array}\right.
\end{eqnarray}
where we used  $R=-4\Lambda$. Considering a static and spherically symmetric spacetime, we can write the line element as 
\begin{eqnarray}
ds^2=e^{2\gamma(r)}dt^2-e^{2\alpha(r)}dr^2-e^{2\beta(r)}\left(d\theta^2+\sin^2\theta d\phi^2\right)\label{ele}\;.
\end{eqnarray}
Thus, the first equation of (\ref{ec}) becomes
\begin{eqnarray}
F^{10}=q\,e^{-(\alpha+\gamma+2\beta)}\;,\;F_{\mu\nu}F^{\mu\nu}=-2q\,e^{-4\beta}\;,\label{em}
\end{eqnarray}
where $q$ is a real integration constant. Substituting  (\ref{em}) into the second equation of (\ref{ec}), we get
\begin{eqnarray}\label{ec1}
\left\{\begin{array}{lll}
\gamma^{\prime\prime}+\gamma^{\prime}\left(\gamma^{\prime}-
\alpha^{\prime}+2\beta^{\prime}\right)&=&e^{2\alpha}
\left(\eta q^2\,e^{-4\beta}-\Lambda\right)\;,\\
\gamma^{\prime\prime}+2\beta^{\prime\prime}+
\gamma^{\prime}\left(\gamma^{\prime}-\alpha^{\prime}
\right)+2\beta^{\prime}\left(\beta^{\prime}-
\alpha^{\prime}\right)&=&e^{2\alpha}
\left(\eta q^2\,e^{-4\beta}-\Lambda\right)\;,\\
e^{2(\alpha-\beta)}-\left[\beta^{\prime\prime}+
\beta^{\prime}\left(\gamma^{\prime}-\alpha^{\prime}+2
\beta^{\prime}\right)\right]&=&e^{2\alpha}
\left(\eta q^2\,e^{-4\beta}+\Lambda\right)\;,
\end{array}\right.
\end{eqnarray}
where $^{\prime}$ denotes the derivative with respect to the radial coordinate $r$. Then, we choose the quasi-global coordinate system 
\begin{eqnarray}
\alpha(r)=-\gamma(r)\;,\;\beta(r)=\ln(r)\;,
\end{eqnarray}
and the second equation of  (\ref{ec1}) becomes 
\begin{eqnarray}
\left(e^{2\gamma}\right)^{\prime\prime}+\frac{2}{r}\left(e^{2\gamma}\right)^{\prime}=2\left(\eta\frac{q^2}{r^4}-\Lambda\right)\;.\label{eq1.1}
\end{eqnarray}
Integrating (\ref{eq1.1}), we obtain the following line element and Maxwell field (or phantom) as solutions of the equations of motion of the action (\ref{action})
\begin{eqnarray}
ds^{2}&=&f(r)dt^2-\left[f(r)\right]^{-1}dr^2-r^2\left(d\theta^2+\sin^2\theta d\phi^2\right)\label{sol}\;,\\
F&=&-\frac{q}{r^2}dr\wedge dt\;,\;f(r)=1-\frac{2M}{r}-\frac{\Lambda}{3}r^2+\eta\frac{q^2}{r^2}\;.\label{max}
\end{eqnarray}
Again, the mass  $M$ and the constant  $1$, of $f(r)$, are determined for satisfying the newtonian limit when $\Lambda=0$, the spacetime being asymptotically that of Minkowski. The solution (\ref{sol}) is that of Reissner-Nordstrom-(A)dS (RN-AdS), for $\eta=1$, and that of anti-Reissner-Nordstrom-(A)dS (phantom or anti-RN-AdS), for $\eta=-1$.
\par
\begin{figure}[h]
\centering
\includegraphics[height=3cm,width=5cm]{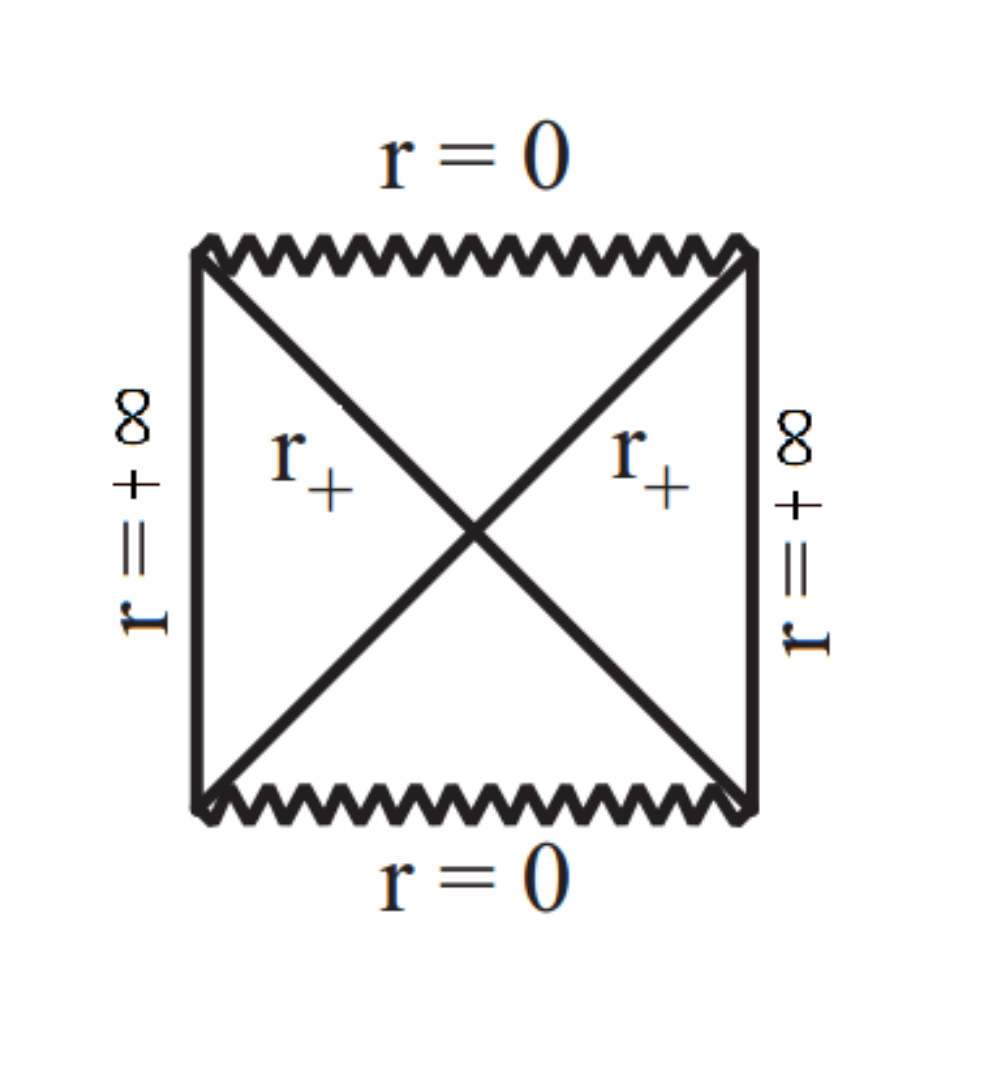}
\caption{\scriptsize{Carter-Penrose diagram for the new anti-RN-AdS solution.}} \label{diagrama}
\end{figure}

The horizons of this solution can be determined by the roots of $f(r)$. We can obtain only two real roots for the equation $f(r)=0$, which are the external horizon $r_+$ (events horizon), or internal horizon $r_-$ , where $0<r_-<r_+$, for $\eta=1$, and $r_-<0<r_+$, for $\eta=-1$. We are just interested to the anti-Reissner-Nordstrom-AdS solution($\Lambda<0$), and to value of $r_+$, which can be represented as \cite{sengupta}
\begin{eqnarray}
r_+&=&\frac{1}{2}\left(\sqrt{x}+\sqrt{\frac{6}{\Lambda}-x-\frac{12M}{\Lambda\sqrt{x}}}\right)\;,\\
x&=&A+B+\frac{2}{\Lambda}\;,\;A=-\sqrt[3]{\frac{2}{y}}\left(\frac{1-4\eta\Lambda q^2}{\Lambda}\right)\;,\;B=-\sqrt[3]{\frac{y}{32}}\left(\frac{3}{\Lambda}\right)\,,\\
y&=&2-36\Lambda M^2+24\eta\Lambda q^2+\sqrt{(2-36\Lambda M^2+24\eta\Lambda q^2)^2-4(1-4\eta\Lambda q^2)^3}\;.
\end{eqnarray} 
The causal structure of the new anti-Reissner-Nordstrom-AdS (anti-RN-AdS) solution, firstly obtained in this work, is identical to the Schwarzschild solution, but with timelike infinite spacial ($r\rightarrow\infty$), as shown in Figure \ref{diagrama}.

The first lozenge  at the right side of the diagram represents a region of space-time with the usual asymptotic region, characteristic of asymptotically AdS spaces. There also exists the possibility to pass through the horizon $ r_+$ to another region, causally disconnected from the first one, and which possesses a spacelike singularity in $r=0$.
\par
The motion of neutral or charged particles can be performed in the same way as in \cite{cruz}-\cite{r4}, following the symmetry $q^2\rightarrow -q^2$, as evinced in the solution (\ref{sol}), which is the crucial difference between the physical interpretations of solutions of RN-AdS and anti-RN-AdS. This aspect deserves to be widely developed and therefore, we intend to address it as the main subject of a future work.
\par
For studying the thermodynamic properties of this solution, it is necessary to express the mass $M$ in terms of the radius of the events horizon $r_+$ and the charge $q$ as follows. Equating $g_{00}=f(r)$ to zero, we get
\begin{eqnarray}
M=\frac{r_+}{2}\left(1-\frac{\Lambda}{3}r_+^2+\eta\frac{q^2}{r_+^2}\right)\label{mass1}\;.
\end{eqnarray}
Now, we are interested to do a geometrical analysis, representing the semi-classical effects of the gravitation
of the black hole solutions previously cited, i.e, quantizing other fields called matter field, treating classically the gravitational field as the background. Then, we will make the semi-classical thermodynamic of black holes, initiated by Hawking \cite{hawking1} and developed posteriorly by other authors \cite{davies2}.\par
There exist various forms of obtaining the Hawking temperature, as for example by the Bogoliubov coefficients \cite{ford}, the metric euclidianization \cite{hawking2}, by the energy-momentum tensor \cite{davies,davies2}, by the transmission and reflection coefficients \cite{gerard3,kanti,k21,r5}, by the analysis of the anomaly term \cite{robinson} and by the superficial gravity of black hole \cite{jacobson}. Until now, these methods appear to be equivalent \cite{manuel}, therefore, we propose to use in this work the method of the calculation of the Hawking temperature through the superficial gravity.\par
The superficial gravity of a black hole is given by the expression \cite{wald}:
\begin{equation}
\kappa =\left[  \frac{g_{00}^{\prime}}{2\sqrt{-g_{00}g_{11}}}\right]_{r\;=\;r_{+}}\, ,\label{10}
\end{equation}
where $r_{+}\;$ is the radius of the events horizon and the Hawking temperature is related with the superficial gravity by the expression \cite{hawking1,jacobson}:
\begin{equation}
T=\frac{\kappa}{2\pi}\, .\label{11}
\end{equation}
Then, for the case of black hole (\ref{sol}), we find that the superficial gravity (\ref{10}) is 
\begin{equation}
 \kappa =\frac{M}{r_+^2}-\frac{\Lambda}{3}r_+-\eta\frac{q^2}{r_+^2}\, ,
\end{equation}
in which, substituting the mass (\ref{mass1}), yields
\begin{equation}
 \kappa =\frac{1}{2r_+}\left(1-\Lambda r_+^2-\eta\frac{q^2}{r_+^2}\right)\,.\label{gsup}
\end{equation}
and the Hawking temperature (\ref{11}), in this case, is :
\begin{equation}
T=\frac{1}{4\pi r_+}\left(1-\Lambda r_+^2-\eta\frac{q^2}{r_+^2}\right)\,.\label{th}
\end{equation}
We define the area of the black hole horizon as 
\begin{equation}
A=\int _{0}^{2\pi}\int_{0}^{\pi}\sqrt{g_{22}g_{33}}\;d\theta d\phi\Big|_{r=r_{+}}=4\pi r^2\Big|_{r=r_+}=4\pi r_{+}^{2}\label{a1}\;. 
\end{equation}
Thus, one can define the entropy of the black hole by \cite{bardeen}
\begin{equation}
S=\frac{1}{4}A=\pi r_{+}^{2}\label{s1}\;. 
\end{equation}
We also calculate, from (\ref{max}), the electric potential at the events horizon as 
\begin{equation}
A_{0}=\int_{+\infty}^{r}F_{10}(r^{\prime})dr^{\prime}\Big|_{r=r_+}=\frac{q}{r_{+}}\label{pot1}\;.
\end{equation}
Through the equations (\ref{mass1}) and (\ref{s1}), we can take the differential of the mass and of the entropy 
\begin{eqnarray}
\label{dif} \left\{\begin{array}{ll}
dM=\frac{1}{2}\left(1-\Lambda r_+^2-\eta\frac{q^2}{r_+^2}\right)dr_+ +\eta\frac{q}{r_+}dq\; ,\\
dS=2\pi r_+ dr_+\;,
\end{array}\right.
\end{eqnarray}
which, with (\ref{th}) and (\ref{pot1}), satisfy the first law of the thermodynamics of black holes \cite{bardeen}
\begin{equation}
dM=TdS+\eta A_{0}dq\label{plt}\;.
\end{equation}
Note that the first law is generalized for the case anti-RN-AdS, with $\eta=-1$, where the sign of the second term, which is related to the work, changes due to the contribution of negative energy to the system.
\par
We can correctly establish the study of the thermodynamics of the anti-RN-AdS system. From the equation (\ref{s1}) we get  $r_+=\sqrt{S/\pi}$, and inserting it in (\ref{mass1}), one gets
\begin{eqnarray}
M(S,q)=\frac{1}{2}\left(\frac{S}{\pi}\right)^{3/2}\left(\frac{\pi}{S}-\frac{\Lambda}{3}+\eta\pi^2\frac{q^2}{S^2}\right)\label{mass2}\;.
\end{eqnarray}
From (\ref{plt}), we obtain the following equations of state 
\begin{eqnarray}
\left(\frac{\partial M}{\partial S}\right)_{q}=T\;,\;\left(\frac{\partial M}{\partial q}\right)_{S}=\eta A_0\;.\label{eqe}
\end{eqnarray}
From the expression (\ref{mass2}), we see that the equations of state (\ref{eqe}) are satisfied. Now, we have all the required ingredients for doing an analysis of the thermodynamic system for the RN-AdS phantom solution. In next section, we will present a brief summary of Geometrotermodynamic method, which will be applied later.


\section{\large The Geometrothermodynamics}

The method of Geometrothermodynamics (GTD uses the differential geometry tool for representing a physical thermodynamic system. With this, we can construct a mathematical space where  all the thermodynamic quantities can be defined, extensive and intensive, also the postulates and equations of state. Let us consider a $(2n+1)-$dimensional space $\mathbb{T}$, whose the coordinates are thermodynamic potential $\Phi$, extensive variables $E^{a}$ and intensive variables $I^{a}$, where $a=1,...,n$. If the space $\mathbb{T}$ possesses a non-degenerated metric  $G_{AB}(Z^{C})$, where $Z^{C}=\{ \Phi , E^{a} , I^{a}\}$, and a Gibbs 1-form $\Theta=d\Phi -\delta_{ab}I^{a}dE^{b}$, where $\delta_{ab}$ is the Kronecker delta, then if the condition  $\Theta\wedge \left(d\Theta\right)^{n}\neq 0$ is satisfied, the set $\left(\mathbb{T} , \Theta , G\right)$ is called contact Riemann manifold \cite{hermann}. The space $\mathbb{T}$ is called thermodynamic phase space. 
\par
One can also define a $n$-dimensional subspace  $\mathbb{E}\subset \mathbb{T}$ with extensive coordinates
 $E^{a}$ and a map  $\varphi :\mathbb{E}\rightarrow\mathbb{T}$, with $\Phi\equiv \Phi (E^{a})$, such that 
\begin{eqnarray}\label{cond1}
\varphi^{*}(\Theta)\equiv 0 \Rightarrow \left\{\begin{array}{ll}
d \Phi=\delta_{ab}I^{a}dE^{b}\;,\\ 
\frac{\partial\Phi}{\partial E^{a}}=\delta_{ab}I^{b}\;. \end{array}\right.
\end{eqnarray}
Then, we call $\mathbb{E}$ by the space of equilibrium thermodynamic states, and the first equation of (\ref{cond1}) by ``first law of thermodynamics", and the second equation by the condition for thermodynamic equilibrium. We impose ``the second law of thermodynamics" as a required condition:
\begin{equation}
\pm\frac{\partial^{2}\Phi}{\partial E^{a}\partial E^{b}}\geq 0\; ,\label{segunda}
\end{equation}
where the sign ($\pm$) depends on the chosen thermodynamic potential. In the mass case we get the sign ($+$) and for the entropy case we get ($-$). This condition is known as convexity of the thermodynamic potential. The pullback  $\varphi^{*}:T^{*}(\mathbb{T})\otimes T^{*}(\mathbb{T})\rightarrow T^{*}(\mathbb{E})\otimes T^{*}(\mathbb{E})$\footnote{Where $T^{*}(\mathbb{T})$ and $T^{*}(\mathbb{E})$ represent the tangent space to $\mathbb{T}$ and $\mathbb{E}$  respectively.}, induces a metric in  $\mathbb{E}$, such that $\varphi^{*}(G)=g$.
\par
Hernando Quevedo has improved probable metric  $G$ for the GTD, such that for the case of black holes, it can be written as follows \cite{quevedo}
\begin{equation}
dL^{2}=G_{AB}dZ^{A}dZ^{B}=\Theta^{2}+\left(\delta_{ab}E^{a}I^{b}\right)\left(\eta_{ab}dE^{a}dI^{b}\right)\; \label{mt}\;,
\end{equation}
where $\eta_{ab} =\{\pm 1,1,...,1\}$, for which the second order phase transition leads to  $\eta_{ab} =\{ -1,1,...,1\}$. Pullback  $\varphi^{*}$ induces a metric in $\mathbb{E}$:
\begin{align}
dl^{2} & =g_{ab}dE^{a}dE^{b}=\frac{\partial Z^{A}}{\partial E^{a}}\frac{\partial Z^{B}}{\partial E^{b}}G_{AB}dE^{a}dE^{b},\nonumber\\
 & =\left( E^{c}\frac{\partial\Phi}{\partial E^{c}}\right) \left( \eta_{ad}\delta^{di}\frac{\partial^{2}\Phi}{\partial E^{i}E^{b}}\right)dE^{a}dE^{b}\; . \label{me}
\end{align}
Since the thermodynamic system does not depend on the choice of the thermodynamic potential and is invariant by the Legendre transformations, we find that the metrics  (\ref{mt}) and  (\ref{me}) are defined in such a way to remain invariants under the Legendre transformations as follows:
\begin{align}
\Phi=\widehat{\Phi}-\delta_{ab}\widehat{E}^{a}\widehat{I}^{b}\; , \; E^{a}=-\widehat{I}^{a}\; , \; I^{b}=\widehat{E}^{b}\; .\label{tl}  
\end{align}
The $n$-dimensional space $\mathbb{E}$, whose the metric is   $g_{ab}$, is used for interpreting the thermodynamic interactions, phase transitions and fluctuations or stability of the thermodynamic system. Through the metric  (\ref{me}), we can calculate the curvature scalar $R$, which provides two interpretations: whether there exist thermodynamic interaction and phase transition, and at which point of the  space of thermodynamic equilibrium states they occur.
\par
We have seen that a metric of the space of thermodynamic equilibrium states $\mathbb{E}$ can be obtained by the pullback of the metric of the contact Riemann space. By the definition of the line element (\ref{me}), of the space  $\mathbb{E}$, one can define the distribution of probability in the interval $E^{a}+dE^{a}$ for finding the physical state with the extensive variable $E^{a}$, by
\begin{align}\label{prob}
P(E^{a})=\frac{\sqrt{\det\left[g_{ab}\right]}}{(2\pi)^{\frac{n}{2}}}\exp\left[\frac{1}{2}g_{ab}dE^{a}dE^{b}\right]\; , 
\end{align}
where $P(E^{a})$ satisfies 
\begin{align}
\int\prod\limits^{n}_{a=1}dE^{a}P(E^{a})=1\; .
\end{align}
By deriving (\ref{prob}) with respect to  $V^{-1}$, where $V$ is the volume of the system, the expression for the second fluctuations can  be obtained
 (in the thermodynamic limit  $V\rightarrow\infty$) \cite{ruppeiner2}
\begin{align}
\langle \Delta E^{a}\Delta E^{b}\rangle =-g^{ab}\; ,\label{flut}
\end{align}
where $\Delta E^{a}=E^{a}-E^{a}_{(0)}$ and  $g ^{ab}$ is the inverse of $g_{ab}$. The unities must be adjusted with a realistic analysis. In the case where the fluctuations are real and small the system is stable.
\par
Another criterion used for determining the stability is through the following objects
\begin{eqnarray}
p_{1}^{(1)}&=&g_{11}>0\; ,\;p_{1}^{(2)}=g_{22}>0\; ,\, ...\, ,\; p_{1}^{(n)}=g_{nn}>0\; , 
\end{eqnarray}
\begin{eqnarray} 
p_{2}^{(1)}&=&\left| \begin{array}{cc}  g_{11} & g_{12} \\
 g_{12} & g_{22}\end{array}\right|>0 \; ,\; p_{2}^{(2)}=\left| \begin{array}{cc}  g_{22} & g_{23} \\
 g_{23} & g_{33}\end{array}\right|>0\; ,\\
p_{3}^{(1)}&=&\left| \begin{array}{ccc}  g_{11} & g_{12}&g_{13} \\
 g_{12} & g_{22}&g_{23}\\
 g_{13} & g_{23}&g_{33}\end{array}\right| >0\; , \; p_{n}=\det\left[ g_{ab}\right]>0\; .
\end{eqnarray}
The positive (negative) sign  of $p_{n}$, which depends on the choice of the thermodynamic potential, gives us the notion of the local stability (instability)  of the thermodynamic system  and the fulfilment of  \cite{bellucci}
\begin{align}
p_{i}>0\; ,\; i=1,\, ...\, ,n\; ,\label{eg}
\end{align}
informs us that the system is globally stable.
\par
Since we are not studying rotating black holes, another mechanism for determining the global stability (instability) is the Helmholtz free energy. In thermodynamic variables of black holes, the Helmholtz free energy is a Legendre transformation of the mass (energy) $M(S,q)$:
\begin{align}
F(T,q)=M(S,q)-TS\;.\label{fh}
\end{align}
When 
\begin{align}
F(E^{a})<0\; ,\; \forall E^{a}\; (E^{a}\in I(E^{a}))\; , \label{egh} 
\end{align}
where $I(E^{a})$ is an interval and  $E^{a}$ are the extensive variables, the thermodynamic system is globally stable. In the usual case of the thermodynamics the  Helmholtz free energy is given by  $F(T,V)=U-TS$.
\par 
We can also define the Gibbs potential as 
\begin{equation}
G(T,A_{0})=M(S,q)-TS-\eta A_{0}q\label{gibbs}\;,
\end{equation}
and the global stability is given by 
\begin{align}
G(E^{a})<0\; ,\; \forall E^{a}\; (E^{a}\in I(E^{a}))\; , \label{eggibbs} 
\end{align}
where we  introduced the sign of $\eta$ in (\ref{gibbs}) for compensating the contribution of the thermodynamic work. Thus, we will use the Gibbs potential for determining the global stability of the thermodynamic system. In the next section, we will explain these methods for the study of the thermodynamic properties of the RN-Ads and anti-RN-Ads solutions. 


\section{\large Termodynamics of phantom Reissner-Nordstrom-AdS black hole}

\subsection{Geometrothemodynamics Application}


We start defining the thermodynamic variables of the system. For the cases of black holes coming from the Einstein-Maxwell theory, we always will get a solution with two physical parameters, the mass $M$ and the charge $q$. Other elements which could be defined from these two parameters are the entropy $S$, the temperature $T$ and the electric potential scalar $A_{0}$. The contact Riemann manifold $\mathbb{T}$  in this case is a $5$-dimensional space and the space of thermodynamic equilibrium states $\mathbb{E}$ is a $2$-dimensional  submanifold.   
\par
The description of the thermodynamics is that of the mass representation $M(S,q)$ in (\ref{mass2}), as thermodynamic potential $\Phi$, defined in the previous section. The extensive variables are the entropy  $S$ and the charge $q$, which was represented by the coordinates $E^{a}$. The intensives variables  are the temperature $T$ and the electric potential $A_{0}$, which was represented by the coordinates $I^{a}$.
\par
Hence, we have the coordinates of the thermodynamic phase space $\mathbb{T}$ as  $Z^{A}=\{ M(S,q), S, q, T, A_{0}\}$, and the Gibbs 1-form is given by \footnote{This expression comes from the first law of thermodynamics  (\ref{plt}).} 
\begin{equation}
\Theta_{M}=dM-TdS-\eta A_{0}dq\;,\label{fgs}
\end{equation}
such that we have $\varphi^{*}(\Theta_{S})=0$, resulting into the first law of the thermodynamics of black holes  $dM=TdS+\eta A_{0}dq$ ($\eta =\pm 1$). 
\par
The line element (\ref{mt}), of the space $\mathbb{T}$, for a  phase transition of second order is given by 
\begin{eqnarray}
dL^{2}&=&\left(dM-TdS-\eta A_{0}dq\right)^{2}+\left(TS+\eta A_{0}q\right)\left[-dSdT+d\left(\eta A_{0}\right)dq\right]\label{mt1}\;.
\end{eqnarray}
The first law, second law and the equations of state at the equilibrium are given by 
\begin{align}
d\Phi & =\delta_{ab}I^{a}dE^{b}\rightarrow dM=TdS+\eta A_{0}dq\;,\\
& \frac{\partial^{2} M}{\partial S^{2}}\,, \frac{\partial^{2} M}{\partial S \partial q}\, , \frac{\partial^{2} M}{\partial q^{2}}\leqslant 0\;,\\
\frac{\partial \Phi}{\partial E^{a}} & =\delta_{ab}I^{b}\rightarrow \frac{\partial M}{\partial S}=T\, ,\, \frac{\partial M}{\partial q}=\eta A_{0}\; .
\end{align}
Now, we must specify the anti-RN-AdS solution(\ref{sol}). The line element (\ref{me}) of the space of thermodynamic equilibrium states, taking into account (\ref{mass2}), is given by 
\begin{eqnarray}
dl^{2}&=&\left(S\frac{\partial M}{\partial S}+q\frac{\partial M}{\partial q}\right)\left(-\frac{\partial^2 M}{\partial S^2}dS^2+\frac{\partial^2 M}{\partial q^2}dq^2\right)\;,\label{me1}\\
&=&g_{SS}dS^2+g_{qq}dq^2\label{marng},\\
g_{SS}&=&\left(\frac{\pi S-\Lambda S^2+3\eta \pi^2 q^2}{4\pi S}\right)\left(\frac{\Lambda S^2+\pi S-3\eta\pi^2 q^2}{8\pi^2 S^2}\right)\;,
\end{eqnarray}
\begin{eqnarray}
g_{qq}&=&\eta\left(\frac{\pi S-\Lambda S^2+3\eta \pi^2 q^2}{4\pi S}\right)\;.
\end{eqnarray} 
We recall here that the phantom contribution may change the signature of the space metric $\mathbb{E}$. For calculating the scalar of curvature associated to the metric (\ref{marng}), we can use the formula of a scalar of a two dimensional space: 
\begin{eqnarray}
R(M,q)&=&-\frac{1}{\sqrt{|det[g]|}}\left[\partial_{q}\left(\frac{\partial_{q}g_{MM}-\partial_{M}g_{Mq}}{\sqrt{|det[g]|}}\right)+\partial_{M}\left(\frac{\partial_{M}g_{qq}-\partial_{q}g_{Mq}}{\sqrt{|det[g]|}}\right)\right]-\frac{det[H_{S}]}{2\left(det[g]\right)^2}\;,\label{ricci}\\
H_{S}&=&\left(\begin{array}{rrr}
g_{MM}&g_{Mq}&g_{qq}\\
\partial_{M}g_{MM}&\partial_{M}g_{Mq}&\partial_{M}g_{qq}\\
\partial_{q}g_{MM}&\partial_{q}g_{Mq}&\partial_{q}g_{qq}
\end{array}\right)\;.\label{harng}
\end{eqnarray}
The scalar (\ref{ricci}) is given by
\begin{eqnarray}
R(S,q)&=&-16\pi^3 S^2\Big( 5\Lambda^3 S^5+\pi\Lambda^2 S^4-2\pi^2\Lambda S^3+18\eta\pi^4 Sq^2+15\eta\Lambda\pi^3 S^2q^2+\nonumber\\
&+&6\eta\Lambda^2\pi^2 S^3q^2-18\pi^5 q^4-63 \Lambda\pi^4 S\Big)\Big/\left(\Lambda S^2+\pi S-3\eta\pi^2 q^2\right)^2\left(\Lambda S^2-\pi S-3\eta\pi^2 q^2\right)^3\,.\label{r1}
\end{eqnarray}
\par
The root of the numerator of (\ref{r1}) can not be found algebraically, therefore,  we perform a numerical analysis here.
\par
By the use of a mathematical software, we find the roots of the scalar of curvature $R(M,q)$ in (\ref{r1}) for the values of the entropy $S_{1}(q=0.25)=0.166306$ and  $S_{2}(q=0.25)=2.76001$ when $\eta=-\Lambda=1$; and $S_{3}(q=0.1)=0.864182$ and  $S_{4}(q=0.1)=2.22539$ when $\eta=\Lambda=-1$. The points where the scalar  $R(M,q)$ diverges are given by the values of the entropy $S_{5}=-(\pi/2\Lambda)(1+\sqrt{1+12\eta \Lambda q^2})$, $S_6=-(\pi/2\Lambda)(-1+\sqrt{1+12\eta\Lambda q^2})$ and $S_7=-(\pi/2\Lambda)(1-\sqrt{1+12\eta\Lambda q^2})$. All the other points are negative or complex values for the entropy and have been rejected. We chose the values of the cosmological constant $\Lambda$ and the electric charge $q$ which are better closed to the most known results in the literature. Here we have basically chosen two values of electric charge, $q=0.1$ and $q=0.25$, and a single value for the cosmological constant $\Lambda=-1$.
\par
For confirming the consistency of our analysis, we can calculate in the usual thermodynamic way the specific heat by the expression
\begin{equation}
C_{q}=\left(\frac{\partial M}{\partial T}\right)_{q}=\left(\frac{\partial M}{\partial S}\right)_{q}\Big/\left(\frac{\partial^2 M}{\partial S^2}\right)_{q}\label{cqs}\;,
\end{equation}    
which leads to 
\begin{eqnarray}
C_q&=&2S\frac{\left(-\pi S+\Lambda S^2+\eta\pi^2 q^2\right)}{\left(\pi S+\Lambda S^2 -3\eta \pi^2 q^2\right)}\;.\label{cqs1}
\end{eqnarray}
The roots of (\ref{cqs1}) give the values where the black hole is extreme, which are $S_8=-(\pi/2\Lambda)(-1+\sqrt{1-4\eta\Lambda q^2})$ and $S_9=(\pi/2\Lambda)(1+\sqrt{1-4\eta\Lambda q^2})<0$, where the last value has been rejected since it is negative ($\Lambda<0$). The values  $S_8$ and $S_9$ can be obtained directly by setting the temperature (\ref{th}) equal to zero and substituting  $r_+=\sqrt{S/\pi}$. The points where the specific heat diverges, i.e., the points of phase transition, are $S_5$ and $S_7$ given previously. More particularly, here, where only the value of $S_5$ is positive for the case anti-RN-AdS, i.e., this case possesses just one point of phase transition. This result is in agreement with that known in the literature, which commonly is demonstrated for the specific heat in terms of the events horizon \cite{niu}, which in our case is a direct substitution of (\ref{s1}). 
\par
\begin{figure}[h]
\centering
\includegraphics[height=3cm,width=5cm]{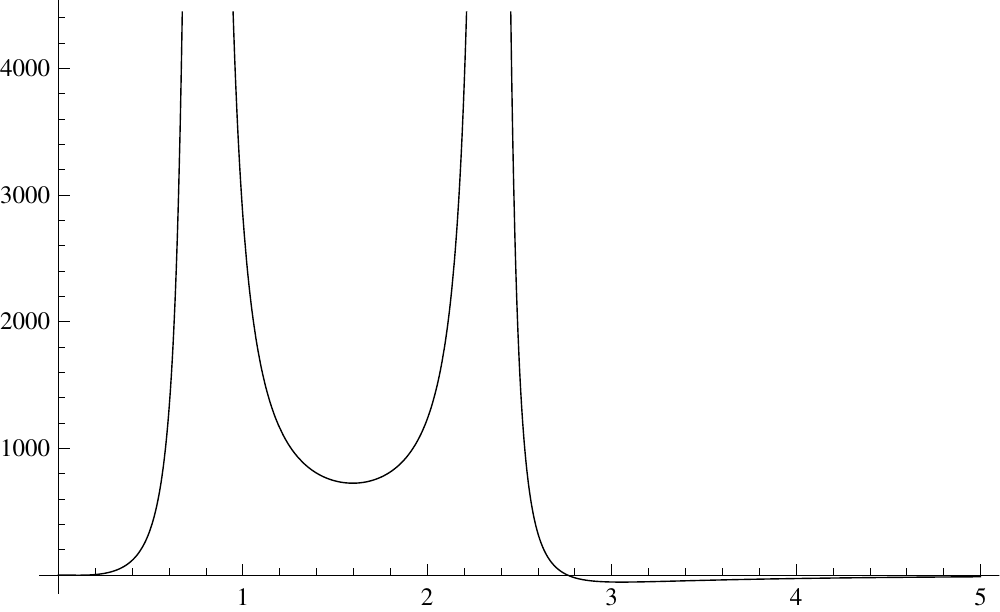}
\caption{\scriptsize{Graph of the curvature scalar in function of the entropy $S$, for the RN-AdS case, with the electric charge  $q=0.25$ and the cosmological constant  $\Lambda=-1$. The main characteristics here are the roots $S_1$ and  $S_1$ and the two points of phase transition $S_5$ and  $S_7$.}} \label{fig1-1}
\end{figure}
\begin{figure}[h]
\centering
\includegraphics[height=3cm,width=5cm]{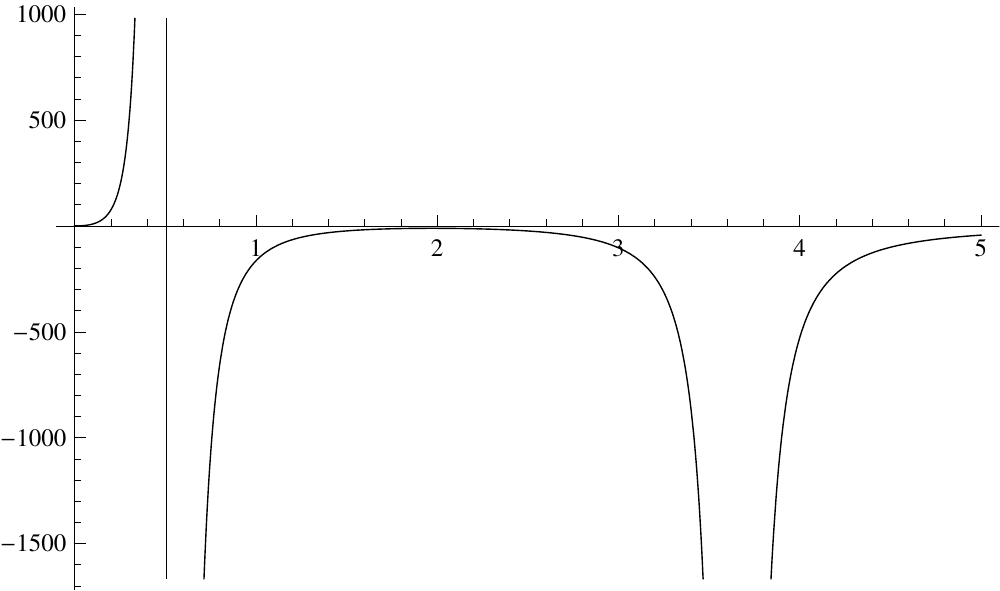}
\caption{\scriptsize{Graph of the specific heat in function of the entropy  $S$, for the RN-AdS case, with the electric charge $q=0.25$ and the cosmological constant $\Lambda=-1$. The main traits here are the two points of phase transition in  $S_5$ and  $S_7$,  an the root in  $S_8$.}} \label{fig1-2}
\end{figure}
\begin{figure}[h]
\centering
\includegraphics[height=3cm,width=5cm]{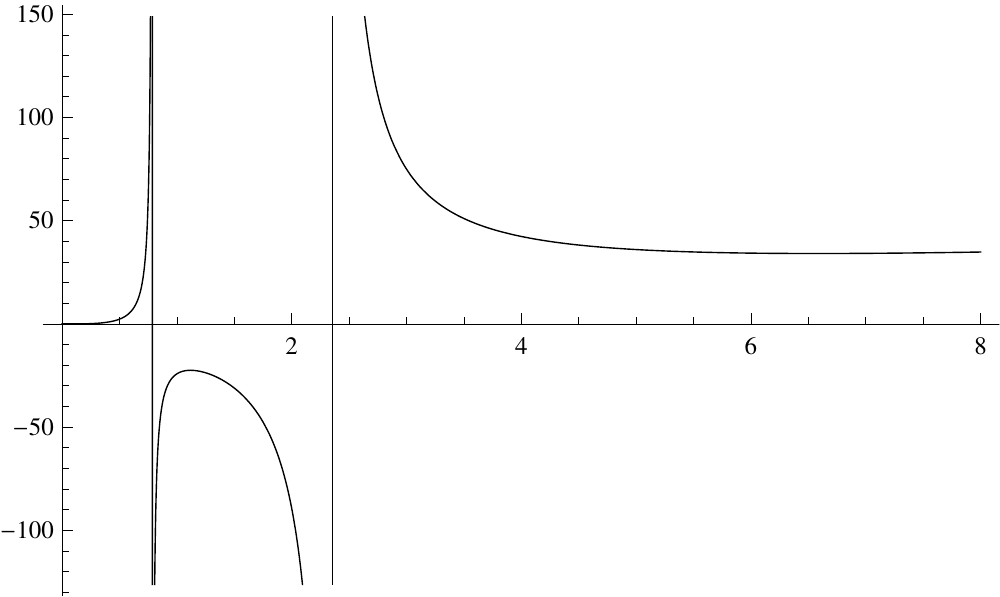}
\caption{\scriptsize{Graph of the curvature scalar in function of the entropy  $S$, for  the anti-RN-AdS case, with the electric charge  $q=0.25$ and the cosmological constant $\Lambda=-1$. The main characteristics here are the two points of phase transition in $S_5$ and $S_6$.}} \label{fig2-1}
\end{figure}
\begin{figure}[h]
\centering
\includegraphics[height=3cm,width=5cm]{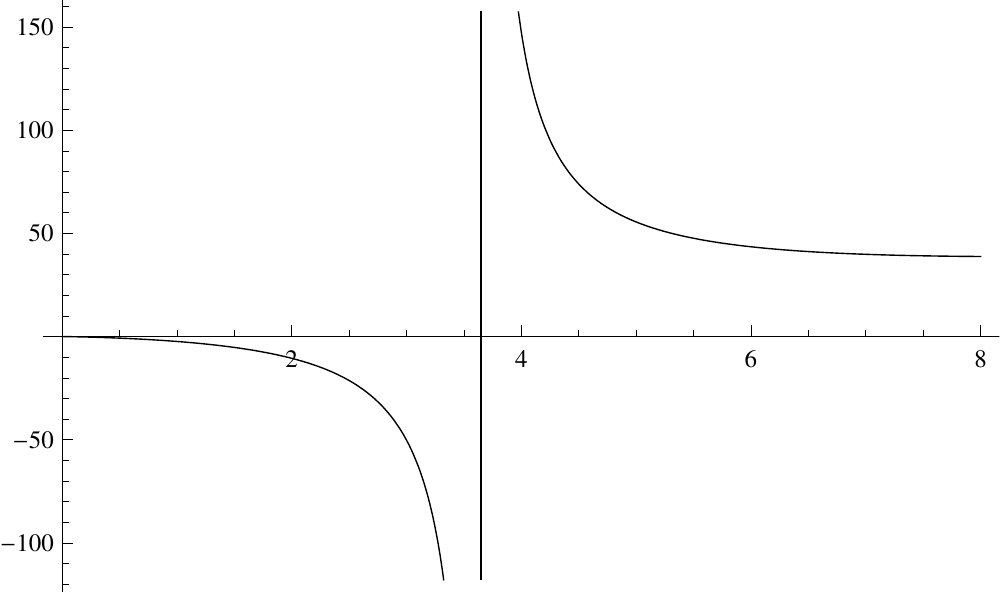}
\caption{\scriptsize{Graph of the specific heat in function of the entropy $S$, for the anti-RN-AdS case, with  the electric charge  $q=0.25$ and the cosmological constant  $\Lambda=-1$. The main feature here is the point of phase transition in  $S_5$.}} \label{fig2-2}
\end{figure}

We present the graphics of the scalar of curvature  (\ref{r1}) and the specific heat (\ref{cqs1}) in function of the entropy, for the fixed value of the electric charge $q=0.25$\footnote{ We plot the graphs with the fixed electric charge $q=0.25$, because they show more clearly the points of phase transition and the roots of the functions.},  in the Figures \ref{fig1-1}-\ref{fig2-2}. For the case RN-AdS, Figures \ref{fig1-1} and \ref{fig1-2}, the unique root for the specific heat is  $S_8(q=0.25)=0.185407$, which differs from the two values which make the scalar of curvature vanishing,  $S_{1}(q=0.25)=0.166306$ and  $S_{2}(q=0.25)=2.76001$. The points of phase transition are identical, located in  $S_5(q=0.25)=2.35619$ and  $S_7(q=0.25)=0.785398$. In the  anti-RN-AdS case, for $q=0.1$, there is any root for which the specific heat vanishes, but there are two roots for which the scalar of curvature vanishes, $S_{3}(q=0.1)=0.864182$ and  $S_{4}(q=0.1)=2.22539$. The figures \ref{fig2-1} and \ref{fig2-2} show that the point of phase transition for the specific heat is in  $S_5(q=0.25)=3.64876$, but for the curvature scalar, there are two points for the phase transition, $S_5(q=0.25)=3.64876$ and  $S_6(q=0.25)=0.507172$, different from the specific heat. Thus, the Geometrothermodynamics method does not provide the same result for the analysis done by the specific heat of the black hole, both for the RN-AdS and for the anti-RN-AdS cases. However, again, the analysis performed by the Geometrothermodynamics method appears to be inefficient, as it has been shown in \cite{zui}.

\subsection{Local and global stability}


The local stability of a thermodynamic system can be studied commonly by the specific heat. But we can also study the components of the metric of the space of thermodynamic equilibrium states $\mathbb{E}$ or the Hessian mass. About the global instability of the system, it can be determined by the analysis of components of the metric $\mathbb{E}$ and all possible determinants of them, by the Helmholtz free energy or by the Gibbs potential. We will study here the local and global stabilities of the class of the black holes solutions of the type RN-AdS and anti-RN-AdS.
\par
Let us start calculating the Hessian of the mass (\ref{mass2}). The Hessian matrix of the mass is defined by 
\begin{eqnarray}\label{HS}
H_{S}=\left(\begin{array}{cc}
\frac{\partial^2 M}{\partial S^2}&\frac{\partial^2 M}{\partial S\partial q}\\
\frac{\partial^2 M}{\partial S\partial q}&\frac{\partial^2 M}{\partial q^2}
\end{array}\right)\;,
\end{eqnarray} 
which for (\ref{mass2}) yields
\begin{eqnarray}\label{HS}
H_{S}=\left(\begin{array}{cc}
\frac{3\eta\pi^2 q^2-\pi S-\Lambda S^2}{8\pi^{3/2}S^{5/2}}&-\eta \frac{q\sqrt{\pi}}{2S^{3/2}}\\
-\eta \frac{q\sqrt{\pi}}{2S^{3/2}}&\eta\sqrt{\frac{\pi}{S}}
\end{array}\right)\;.
\end{eqnarray} 
The components of the Hessian matrix assume clearly the positive and negative values, which implies the local instability for the two solutions, RN-AdS and anti-RN-AdS. 
\par
\begin{figure}[h]
\centering
\includegraphics[height=3cm,width=5cm]{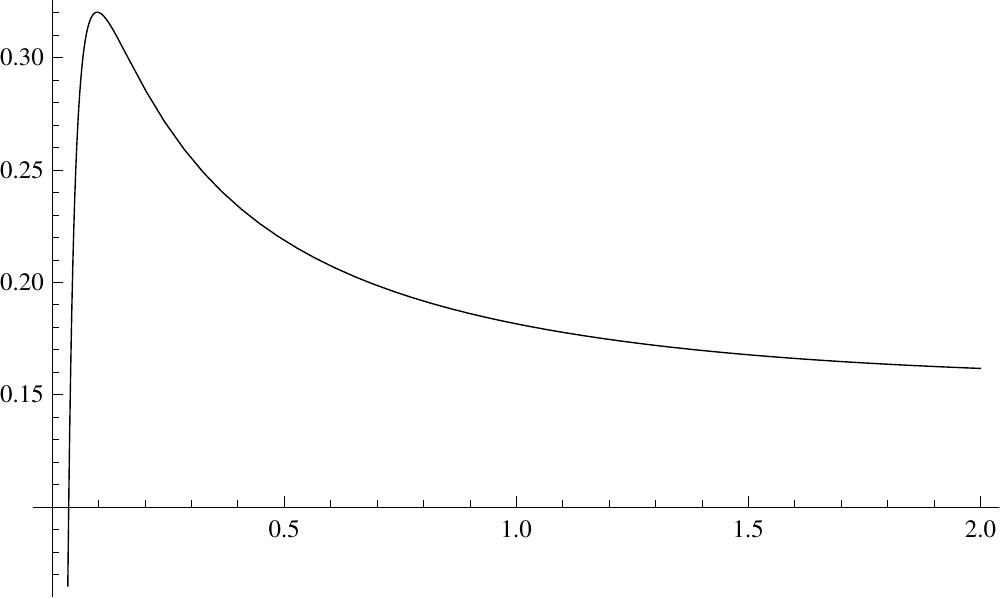}
\caption{\scriptsize{Graph of the temperature in function of the entropy  $S$, for the RN-AdS case, with the electric charge $q=0.1$ and the cosmological constant $\Lambda=-1$.}} \label{temperatura-1}
\end{figure}
\begin{figure}[h]
\centering
\includegraphics[height=3cm,width=5cm]{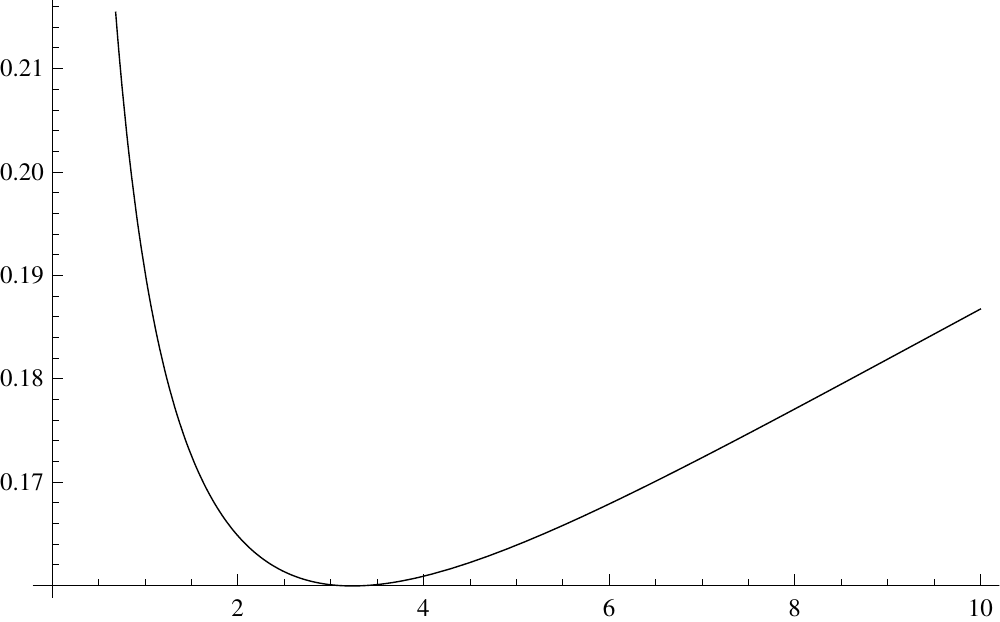}
\caption{\scriptsize{Graph of the temperature in function of the entropy  $S$, for the anti-RN-AdS case, with  the electric charge $q=0.1$ and the cosmological constant $\Lambda=-1$.}} \label{temperatura-2}
\end{figure}

We can also analyse the specific heat (\ref{cqs1}), but to do this, let us first study the temperature of the black hole (\ref{th}). The temperature can be written in terms of the entropy as follows
\begin{eqnarray}
T(S,q)=\frac{\pi S-\Lambda S^2-\eta\pi^2 q^2}{4(\pi S)^{3/2}}\label{t2}\;.
\end{eqnarray}
In the RN-AdS case the temperature is positive for  $S>S_e$, where the value for the extreme case is given by  $S_e=(-\pi/2\Lambda)(-1+\sqrt{1-4\Lambda q^2})$, zero for  $S=S_e$ and negative for $S<S_e$ (unstable), which is in accordance with \cite{rabin2,sengupta}. For the anti-RN-AdS case, the temperature is always positive for $S>0$, then there isn't the extreme case, as we previously mentioned. The graphics of the Figures \ref{temperatura-1} and \ref{temperatura-2} are of temperature in function of the entropy for the two cases. In the RN-AdS case, it is clear that $T=0$ for $S_e$, passing later by the maximum and decreasing slowly. For the anti-RN-AdS case, the temperature starts very high, decreasing until a minimum and later increases more slowly.
\par    
Now, using (\ref{t2}) we can write the specific heat  (\ref{cqs1}) as follows   
\begin{eqnarray}
C_q=\frac{8T(S,q)\pi^{3/2}S^{5/2}}{3\eta\pi^2 q^2-\pi S -\Lambda S^2 }\label{cqs2}\;.
\end{eqnarray}
Analysing the sign of the specific heat for the RN-AdS case, we have a locally unstable phase where $C_q<0$, for $S\in (S_7,S_5)$, where $S_{5,7}=-(\pi/2\Lambda)(1\pm\sqrt{1+12\eta \Lambda q^2})$ and $S_e<S_7$. The black hole is extreme in $S=S_e$ and one has a local stable phase where $C_q>0$, for $S\in(S_e,S_7)$ or $S>S_5$, with $S_e<S_7$, according to the graphic of the Figure \ref{fig1-2}. This also provides a value of the extreme charge, to the RN-AdS case, which is given by $q_e=1/2\sqrt{3}\approx 0.28867513459$ \cite{rabin2}. The anti-RN-AdS case, the graphic of the Figure \ref{fig2-2} shows that there is no extreme case and the system is locally stable for the condition $S>S_5$, since $Cq<0$ for $0<S<S_5$.
\par
We can also study the local and global stabilities by the metric components of the space of thermodynamic equilibrium states $\mathbb{E}$ in (\ref{me1}), as shown in the section $3$. The sign of the metric components (\ref{me1}) for the RN-AdS case are given by $g_{SS}>0$ for $ S\in(S_7,S_5)$, $g_{qq}>0$ for $S>0$ and the determinant of the metric $g>0$ for $S\in(S_7,S_5)$. Collecting the results, we observe that the system is locally and globally stable if $S\in (S_7,S_5)$. This is not in agreement with the analysis made by the specific heat. But for the anti-RN-AdS case, the components $g_{SS}$, $g_{qq}$ and the determinant $g$ are always positive in disjunct intervals. Thus, there does not exist interval for which the system can be locally or globally stable, contradicting again the result of the specific heat method.
\par
\begin{figure}[h]
\centering
\includegraphics[height=3cm,width=5cm]{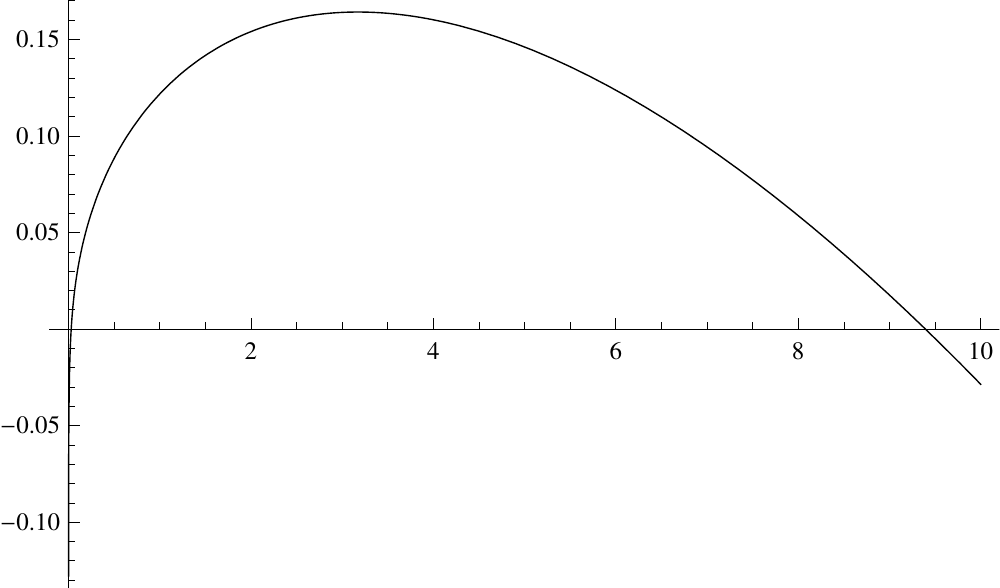}
\caption{\scriptsize{Graph of the Gibbs potential, for the RN-AdS case, in function of the entropy $S$ with the electric charge $q=0.1$ and the cosmological constant  $\Lambda=-1$.}} \label{gibbs-1}
\end{figure}
\begin{figure}[h]
\centering
\includegraphics[height=3cm,width=5cm]{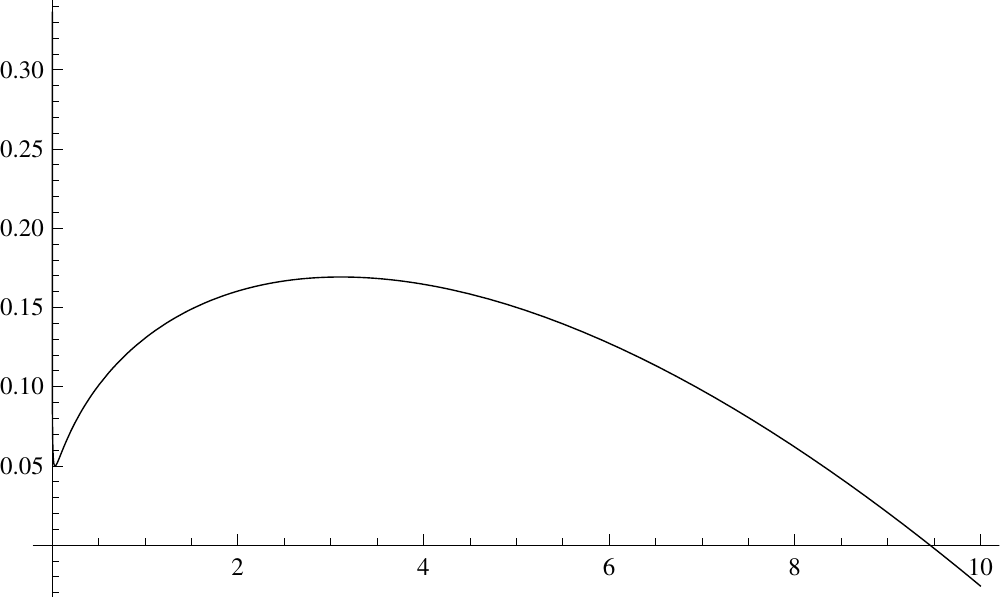}
\caption{\scriptsize{Graph of the Gibbs potential, for the anti-RN-AdS case, in function of the entropy  $S$ with the electric charge  $q=0.1$ and the cosmological constant  $\Lambda=-1$.}} \label{gibbs-2}
\end{figure}

Let us now study the global stability of these solutions. We can rewrite the mass and the electric potential in terms of the entropy and of the electric charge as follows 
\begin{eqnarray}
M&=&\frac{-\Lambda S^2+3\pi S+3\eta\pi^2 q^2}{6\pi^{3/2}\sqrt{S}}\;,\;A_{0}=q\sqrt{\frac{\pi}{S}}\label{eq1}\;.
\end{eqnarray}
Our analysis starts with the grand-canonical ensemble. Using  (\ref{t2}) and (\ref{eq1}), one can calculate the Gibbs potential (\ref{gibbs})
\begin{equation}
G=\frac{\Lambda S^2+3\pi S-3\eta\pi^2 q^2}{12\pi^{3/2}\sqrt{S}}\label{gibbs1}\;.
\end{equation}
Equating (\ref{gibbs1}) with zero, one obtains  
\begin{eqnarray}
S_{G1,G2}=\left(-\frac{\pi}{2\Lambda}\right)\left(3\mp \sqrt{9+12\eta\Lambda q^2}\right)\label{g1}\;.
\end{eqnarray}
In the RN-AdS case, from the Gibbs potential, we get  $G<0$, which is a stable system for $S\in(0,S_{G1})\cup(S_{G2},+\infty)$, and this is in accordance with \cite{sengupta}. In the anti-RN-AdS case, one gets  $G<0$, which is a stable system only when $S>S_{G2}$. The graphs of the Gibbs potential for the two cases are presented in the Figures  \ref{gibbs-1} and \ref{gibbs-2}.
\par
\begin{figure}[h]
\centering
\includegraphics[height=3cm,width=5cm]{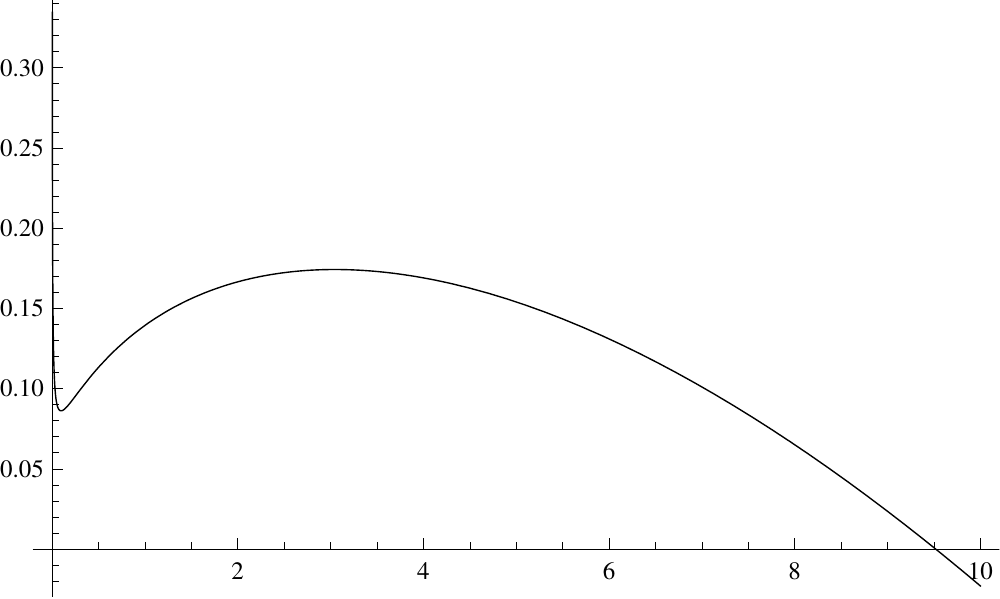}
\caption{\scriptsize{Graph of the Helmholtz free energy, for the RN-AdS case, in function of the entropy $S$ with the value of the electric charge $q=0.1$ and the cosmological constant  $\Lambda=-1$.}} \label{h-1}
\end{figure}
\begin{figure}[h]
\centering
\includegraphics[height=3cm,width=5cm]{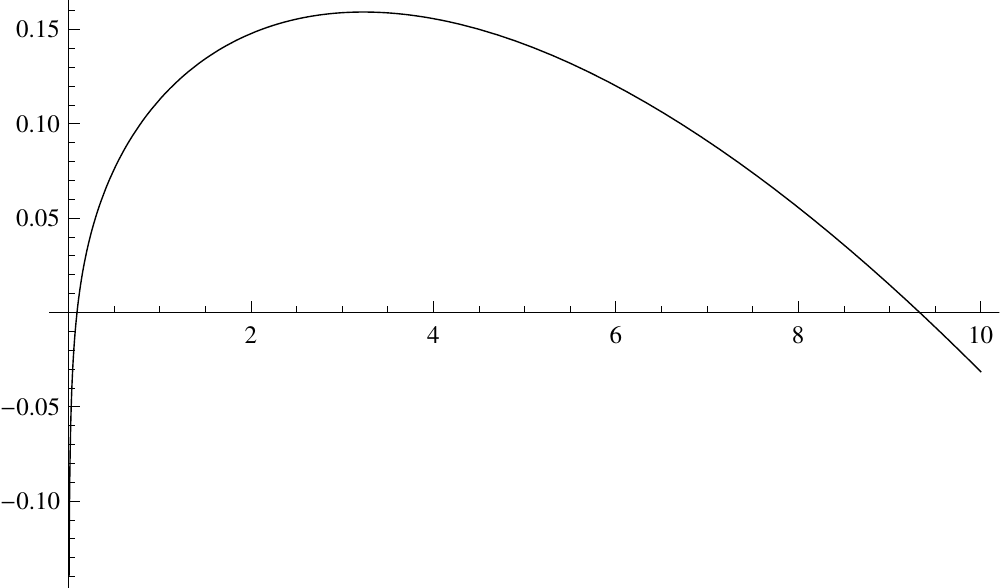}
\caption{\scriptsize{Graph of the Helmholtz free energy, for the anti-RN-AdS case, in function of the entropy $S$ with the values of the electric charge $q=0.1$ and the cosmological constant $\Lambda=-1$.}} \label{h-2}
\end{figure}

We can now analyse the canonical ensemble. The Helmholtz free energy (\ref{fh}), from (\ref{t2}) and (\ref{eq1}), is given by
\begin{eqnarray}
F=\frac{\Lambda S^2+3\pi S+9\eta\pi^2 q^2}{12\pi^{3/2}\sqrt{S}}\label{fh1}\,,
 \end{eqnarray} 
which, equating to zero yields
\begin{eqnarray}
S_{F1,F2}=\left(-\frac{3\pi}{2\Lambda}\right)\left(1\pm \sqrt{1-4\eta\Lambda q^2}\right)\label{g1}\;.
\end{eqnarray}
Now, through the canonical ensemble it seems that we get a inversion between the two cases, the normal and phantom. In the RN-AdS case the system is stable, i.e, $F<0$ for $S>S_{F1}$. In the anti-RN-AdS case the system is stable for $S\in(0,S_{F2})\cup(S_{F1},+\infty)$. The graph of the Helmholtz free energy, for the two cases, are represented in the Figures  \ref{h-1} and \ref{h-2}.
\section{Conclusions}
We obtained a new solution of Einstein-anti-Maxwell theory with cosmological constant, called anti-RN-(A)dS solution. We analysed the basic properties of this solution, such as the horizons. Later, we focused our attention on the thermodynamics of the anti-RN-AdS solution.
\par

We made a summary of the Geometrothermodynamic method. We applied the method to the solutions of types RN-AdS  and anti-RN-AdS, as well as the most usual methods of black holes thermodynamics. Considering the mass as the thermodynamic potential, we obtained the scalar curvature (\ref{r1}) of the space $\mathbb{E}$. This scalar possesses  two roots and two points of phase transition in each case. But these roots are not obtained by the specific heat method. For the RN-AdS case, the unique root for the specific heat is $S_8(q=0.25)=0.185407$, which differs from the two values which cancel the scalar curvature $S_{1}(q=0.25)=0.166306$ and  $S_{2}(q=0.25)=2.76001$. In anti-RN-AdS case, there is any root for the specific heat, but there is two roots of the curvature, $S_{3}(q=0.1)=0.864182$ and  $S_{4}(q=0.1)=2.22539$. So, the Geometrothermodynamic method  does not give us the same result of the analysis made by the specific heat of the black hole, in both cases, i.e, RN-AdS and anti-RN-AdS. Again, the analysis made by the Geometrothermodynamic method  has been ineffective in some way, as in \cite{zui}.
\par
Then we analyse the local and global stability of these solutions, through the Hessian of the mass, of the components and the determinant of the metric of the space of equilibrium thermodynamic states $\mathbb{E}$, the specific heat, the Gibbs potential and Helmholtz free energy. The local stability can not be established by the Hessian of the mass. For a given interval of real values of the entropy, the local stability can be established for the two solutions, using the specific heat. Using the components and  determinant of the metric in $ \mathbb{E}$ we do not obtain a result consistent with to calculus by the specific heat. Finally, through the grand-canonical and canonical ensembles, we established the global stability of the thermodynamic system of the two  RN-AdS and anti-RN-AdS solutions. These ensembles seem to produce results symmetric with each other.
\par 
Therefore, we showed the consistent construction of the thermodynamics of the anti-RN-AdS solution and its global stability. This shows clearly that this solution is physically stable. We expect that this analysis of the thermodynamics system made in this paper leads to a constraint to the spacetime geometry stability of this solution.  
\par
About the Geometrothermodynamics method, we  hope that establishing a new metric of the thermodynamic space $\mathbb{T}$, we can interpret the results more consistently. We believe that the introduction {\it ad hoc} of the  Lorentzian signature (\ref{mt}),  may have a more physical justification. In fact, when we introduce this metric  we hare to impose the result that we expect to thermodynamic system. For example, when  we introduced the Lorentzian metric, we imposed to the thermodynamic system a possible second order phase transition. It will be necessary a greater understanding of the geometry of contact Riemann manifold, such that we can introduce a metric arising from previously established physical requirements, but which may offer posteriorly any known results.
\par
{\bf Acknowledgement}: MER thanks UFES and UFVJM for the hospitality during the elaboration of this work. MJSH thanks CNPq/FAPES for the ﬁnancial support. DFJ thanks UFVJM for the hospitality during the elaboration of this work.



\begin{thebibliography}{10}
\bibitem{hawking1}S. Hawking, Commun. Math. Phys. {\bf 43}, 199 (1975).
\bibitem{bardeen}J. M. Bardeen, B. Carter and S. W. Hawking, Commun. Math. Phys., 31, 161 (1973).
\bibitem{lousto}C. O. Lousto, Nucl.Phys. B {\bf 410}: 155-172 (1993); Erratum-ibid. {\bf 449}: 433 (1995). 
\bibitem{davies}Proc.Roy.Soc.Lond. A {\bf 353}: 499-521 (1977).
\bibitem{quevedo}H. Quevedo, A. Sanchez, S. Taj and A. Vazquez, Gen. Rel. Grav. {\bf 43}:1153-1165 (2011).
\bibitem{louko}Jorma Louko and Stephen N. Winters-Hilt, Phys. Rev. D {\bf 54}:2647-2663 (1996).
\bibitem{hannestad}S. Hannestad, Int. J. Mod. Phys. A {\bf 21}, 1938 (2006);
\bibitem{r1}J. Dunkley et al., Astrophys. J. Suppl. Ser. {\bf 180}, 306 (2009).
\bibitem{kirill1}K.A. Bronnikov, M.S. Chernakova, J.C. Fabris, N. Pinto-Neto and M.E. Rodrigues, Int.J.Mod.Phys.D {\bf 17}:25-42 (2008).
\bibitem{gibbons}G.W. Gibbons and D. A. Rasheed, Nucl. Phys. B {\bf 476}, 515 (1996).
\bibitem{gao}C. J. Gao and S. N. Zhang, arXiv:hep-th/0604114.
\bibitem{grojean}C. Grojean, F. Quevedo, G. Tasinato, and I. Zavala, J. High Energy Phys. {\bf 08} (2001) 005.
\bibitem{hull}C. M. Hull, JHEP {\bf 9807} (1998) 021.
\bibitem{gerard2}Mustapha Azreg-Aïnou, Gérard Clément, Júlio C. Fabris, Manuel E. Rodrigues, Phys.Rev.D {\bf 83}:124001 (2011).
\bibitem{rao}C. R. Rao, Bull. Calcutta Math. Soc. {\bf 37}, 81 (1945). 
\bibitem{amari} S. Amari, Differential-Geometrical methods in Statistics (Springer-Verlag, Berlin, 1985).
\bibitem{weinhold}F. Weinhold, J. Chem. Phys. {\bf 63}, 2479, 2484, 2488, 2496 (1975); {\bf 65}, 558 (1976).
\bibitem{ruppeiner}G. Ruppeiner, Phys. Rev. A {\bf 20}, 1608 (1979).
\bibitem{r2}G. Ruppeiner, Rev. Mod. Phys. {\bf 67}, 605 (1995); {\bf 68}, 313 (1996).
\bibitem{quevedo1} Danny Birmingham, Susan Mokhtari, Phys.Lett.B {\bf 697}:80-84 (2011), arXiv:1011.6654v1 [hep-th].
\bibitem{k1} Hernando Quevedo, Maria N. Quevedo, arXiv:1111.5056v1 [math-ph].
\bibitem{k2} Hernando Quevedo, Alberto Sanchez, Phys.Rev.D {\bf 79}:024012 (2009), arXiv:0811.2524v1 [gr-qc].
\bibitem{k3} Alexis Larranaga, Alejandro Cardenas, arXiv:1108.2205v1 [gr-qc].
\bibitem{k4} Hernando Quevedo, Alberto Sanchez, Safia Taj, Alejandro Vazquez, arXiv:1101.4494v1 [hep-th].
\bibitem{k5} Alexis Larranaga, Sindy Mojica, Brazilian Journal of Physics {\bf 41}: 154-158 (2011), arXiv:1012.2070v1 [gr-qc].
\bibitem{k6} Hernando Quevedo, Alberto Sanchez, JHEP {\bf 0809}:034 (2008), arXiv:0805.3003v2 [hep-th].
\bibitem{k7} Hernando Quevedo, Alberto Sanchez, Phys. Rev. D {\bf 79}, 087504 (2009), arXiv:0902.4488v2 [gr-qc].
\bibitem{k8} Safia Taj, Hernando Quevedo, arXiv:1104.3195v1 [math-ph].
\bibitem{k9} W. Janke, D. A. Johnston, R. Kenna, J. Phys. A {\bf 43}:425206, 2010, arXiv:1005.3392v2 [hep-th].
\bibitem{k10} László Árpád Gergely, Narit Pidokrajt, Sergei Winitzki, Eur.Phys.J.C {\bf 71}:1569 (2011), arXiv:0811.1548v3 [gr-qc].
\bibitem{k11} Hernando Quevedo, J.Math.Phys. {\bf 48} (2007) 013506, arXiv:0604164v2 [physics.chem-ph].
\bibitem{k12} Hernando Quevedo, Alberto Sanchez, Alejandro Vazquez, arXiv:0811.0222v1 [math-ph].
\bibitem{k13} H. Quevedo, A. Sanchez, S. Taj, A. Vazquez, Gen.Rel.Grav. {\bf 43}:1153-1165,2011, arXiv:1010.5599v1 [gr-qc].
\bibitem{k14} Hernando Quevedo, Alejandro Vazquez, AIP Conf.Proc. {\bf 977}:165-172 (2008), arXiv:0712.0868v1 [math-ph].
\bibitem{k15} M. Akbar, H. Quevedo, K. Saifullah, A. Sanchez, S. Taj, Phys.Rev.D {\bf 83}: 084031 (2011), arXiv:1101.2722v1 [gr-qc].
\bibitem{k16} H. Quevedo, A. Sanchez, A. Vazquez, arXiv:0805.4819v5 [hep-th].
\bibitem{k17} Alejandro Vazquez, Hernando Quevedo, Alberto Sanchez, J.Geom.Phys. {\bf 60}: 1942-1949 (2010), arXiv:1101.3359v1 [math-ph].
\bibitem{r3} J.L. Alvarez, H. Quevedo, A. Sanchez, Phys.Rev.D {\bf 77}: 084004 (2008),arXiv:0801.2279v1 [gr-qc]. 
\bibitem{zui}Manuel E. Rodrigues and Zui A. A. Oporto, Thermodynamics of phantom black holes
in Einstein-Maxwell-Dilaton theory, arXiv:1201.5337 [gr-qc].
\bibitem{sengupta}Ratna Koley, Joydip Mitra, Supratik Pal and Soumitra SenGupta, A critical analysis of thermodynamic properties of braneworld black holes in anti-de Sitter spacetime, arXiv:0910.5096v1 [hep-th].

\bibitem{cruz}M. Olivares, J. Saavedra, C. Leiva and J.R. Villanueva, Mod. Phys. Lett. A, Vol. {\bf 26}, No. 39 (2011) pp. 2923-2950, arXiv:1101.0748v2 [gr-qc].
\bibitem{k18} Norman Cruz, Marco Olivares, Joel Saavedra and J.R. Villanueva, arXiv:1111.0924v1 [gr-qc].
\bibitem{k19} Changchun Zhong and Sijie Gao, JETP Letters, Vol. {\bf 94}, No. 8, pp. 589-592 (2011), arXiv:1109.0772v4 [hep-th].
\bibitem{k20} D. Pugliese, H. Quevedo and R. Ruffini, Phys. Rev. D {\bf 83}:104052 (2011), arXiv:1103.1807v3 [gr-qc].
\bibitem{r4}D. Pugliese, H. Quevedo and R. Ruffini,  Phys. Rev. D {\bf 83}:024021 (2011), arXiv:1012.5411v1 [astro-ph.HE].

\bibitem{davies2}N. D. Birrell and P. C. W. Davies, Quantum fields in curved space, Cambridge University Press, 1982.
\bibitem{ford} L. H. Ford. arXiv: gr-qc/9707062.
\bibitem{hawking2} G. W. Gibbons and S. Hawking, Phys. Rev. D {\bf 15}: 2752-2756 (1977).
\bibitem{gerard3}G. Clement, J. C. Fabris and G. T. Marques, Phys. Lett. B {\bf 651}: 54-57 (2007).
\bibitem{kanti}Panagiota Kanti and John March-Russell, Phys.Rev.D {\bf 66}: 024023 (2002).
\bibitem{k21} Wontae Kim and John J. Oh, J.Korean Phys.Soc. {\bf 52}: 986 (2008)
\bibitem{r5} Kazuo Ghoroku, Arne L. Larsen, Phys.Lett. B {\bf 328}: 28-35 (1994).
\bibitem{robinson}S.P. Robinson and F. Wilczek, Phys. Rev. Lett. {\bf 95}: 011303 (2005).
\bibitem{jacobson}T. Jacobson and G. Kang, Class.Quant.Grav. {\bf 10}:L201-L206 (1993); arXiv: gr-qc/9307002.
\bibitem{manuel}Glauber Tadaiesky Marques and Manuel E. Rodrigues, Eur. Phys. J. C 72: 1891 (2012), arXiv:1110.0079v2 [gr-qc].
\bibitem{wald}R. M. Wald, General Relativity, Univ. of Chicago Press, 1984, Chicago.
\bibitem{hermann}R. Hermann, Geometry, physics and systems (Marcel Dekker, New York, 1973); G. Hernandez and E. A. Lacomba, Contact Riemannian geometry and thermodynamics, Diff. Geom. and Appl. 8, 205 (1998).
\bibitem{ruppeiner2}George Ruppeiner, Phys. Rev. D {\bf 75}, 024037 (2007);  Rev. Mod. Phys. {\bf 67}: 605-659 (1995), Erratum-ibid. {\bf 68}: 313-313 (1996).
\bibitem{bellucci}Stefano Bellucci and Bhupendra Nath Tiwari, arXiv:1103.2064v1 [hep-th].
\bibitem{niu}Chao Niu, Yu Tian and Xiaoning Wu, Phys.Rev. D {\bf 85}: 024017 (2012).
\bibitem{rabin2}Rabin Banerjee, Sumit Ghosh and Dibakar Roychowdhury, Phys.Lett. B {\bf 696}:  156-162 (2011).
\bibitem{sengupta}Anurag Sahay, Tapobrata Sarkar and Gautam Sengupta, JHEP {\bf 1007}:  082 (2010). 
\end{thebibliography}
\end{document}